# Building Healthcare – Patient Relationship with CRM 2.0
## Lesson Learnt from Prita Mulyasari's Case

Muhammad Anshari [1] and Mohammad Nabil Almunawar [2]

Faculty of Business, Economic & Policy Studies
Universiti Brunei Darussalam
Brunei Darussalam

[1]anshari@yahoo.com
[2]nabil.almunawar@ubd.edu.bn

*Abstract*— **Healthcare is implementing CRM as a strategy for managing interactions and communication with patients which involves using Information and Communication Technology (ICT) to organize, automate, and coordinate business processes. CRM with the Web technology provides healthcare the ability to broaden service beyond its usual practices, and thus provides a particular advantageous environment for them that want to use ICT to achieve complex healthcare goal. This paper we will discuss and demonstrate how a new approach in CRM will help the healthcare increasing their customer support, and promoting better health to patient. The patients benefited from the customized personal service so that they have full information access to perform self managed their own health and the healthcare provider will have a loyal and retains the right customer. A conceptual framework of approach will be highlighted. Customer centric paradigm in social network's era and value creation of healthcare's business process will be taken into consideration.**

I. INTRODUCTION

Customer Relationship Management (CRM) is how to attract new customers coming to an organization, retaining them throughout the entire lifetime of a relationship, and extending other services or products to the existing customers. Likewise, in the healthcare environment, healthcare providers are challenged to acquiring potential customers for the healthcare services, retaining them to use the services, and extending various services in the future. In order to achieve those strategies, healthcare provider must consider establishing closeness of relationship between patient-healthcare providers, offer convenience of services, and build trust of information sharing.

As a business, healthcare organization stands in need of the same standards of customer service as other business organizations. The fact that customer service expectations in healthcare organization are high poses a serious challenge for healthcare providers as they have to make exceptional impression on every customer. In the competitive commercial healthcare market, poor service and distrusted service leads customers to switch healthcare providers because poor service indicates inefficiency, higher cost and lower quality of care.

Nowadays, more patients have more choices in where they seek care and how they interact with their healthcare providers. A great customer service can lead to major improvements in the health care system. Customer service is not an "extra"— it is an essential requirement for providing high quality healthcare and for staying in highly competitive business (Stanton, 2009). Patients are making clear choices about where they receive care based on service experiences and it is crucial for organizations to create an institutional ability to sense and respond empathetically (Katzenbach Partners, 2008).

Healthcare organization strategies should transform customer strategies and systems to customer engagement. The one is more focused on the conversation that is going on between organization and customer. Conversation between healthcare organization and customers is important to initiate trust between both parties. Proactive strategies will improve customer services. And great customer support will increase loyalty, revenue, brand recognition, and business opportunity.

This paper is partially motivated by the case between a patient whose name is Prita Mulyasari (Prita) with the Omni International Hospital Tangerang, Indonesia (Omni). The case generated massive public attention channels through various media, including social network sites.

*Prita Mulyasari, house mother of two children, for allegedly defaming a hospital via an online complaint, triggered unprecedented public protest across the country. The case started when she wrote an email to her friends in September 2008 detailing her complaint and dissatisfaction towards her experiences at Omni International Hospital Alam Sutera Tangerang; she felt a poor treatment, wrong diagnoses, physician who always late in visiting the patient, and inaccessibility of medical records prompted her dissatisfaction to the hospital's service, which was soon the email rapidly distributed across forums via online mailing lists. Once the email became public knowledge, Omni responded by filing a criminal complaint and a civil lawsuit against Prita.*

*Then, verdict against Prita, at Banten District Court on May 13, 2009, she was sentenced to six years jail and fines for defamation and sending complaint email publicly. The case generated massive public attention, rallies were held across the country. Virtual community supported through Facebook by collecting coins to support Prita paying her fines to the Hospital. While significant pressure eventually led to Prita being released from detention on June 3, 2009, and ordered to remain under city arrest due to humanitarian reasons. (Detik.com, 2008; Caveat, 2009)*

We will use this case to deploy a CRM model to address relationship management between customer (patient) and healthcare organizations that incorporate the latest development in Information and Communication Technology (ICT).

## II. BACKGROUND OF STUDY

*Customer Relationship Management*

CRM is a broad term and widely-implemented strategy for managing interactions with customers which involves using technology to organize, automate, and synchronize business processes—principally customer service, marketing, and sales activities. The overall goals are to find, attract, and win new customers, nurture and retain those the company already has, entice former customers back into the fold, and reduce the costs of marketing and customer service (Gartner, 2009).

Greenberg (2009) defined CRM is a philosophy and a business strategy supported by a system and a technology designed to improve human interactions in a business environment. Furthermore, it is an operational, transactional approach to customer management that is focused around the customer facing departments, sales, marketing and customer service. Furthermore, the early CRM initiatives was the process for modification, culture change, technology and automation through use of data for support the management of customers so it can meet a business value of corporate objectives such as increases in revenue, higher margins, increase in selling time, campaign effectiveness, reduction in call queuing time, etc.

Although the development of CRM has been mature, there are many challenges in adopting CRM for healthcare organizations. Due to the complexity of the business nature in healthcare there are many issues dealing with patients must be considered. A healthcare is undergoing a paradigm shift, from 'Industrial Age Medicine to Information Age Healthcare' (R. Smith, 1997). This 'paradigm shift' is shaping health systems (Haux et.al, 2002). It is also transforming the healthcare-patient relationship (Ball, 2001). For example, World Wide Web has changed the way the public engage with health information (Powell et al., 2003). According to Pew Internet and American Life Project, large shares of Internet users say that they will first use Internet when they need Information about healthcare (Pew Internet, 2005). People begin to use Internet resources for research on the health information and services that they are interested in using. ICT creates an environment where patients can explore clinical records and health education programs at a time.

*CRM 2.0*

Figure 1 shows the business strategy on CRM. The model is a hybrid and having three key phases and three contextual factors; three key phases are customer acquisition, retention, and extension. And the other three contextual factors are marketing orientation, value creation, and innovative IT (Marketingteacher, 2010).

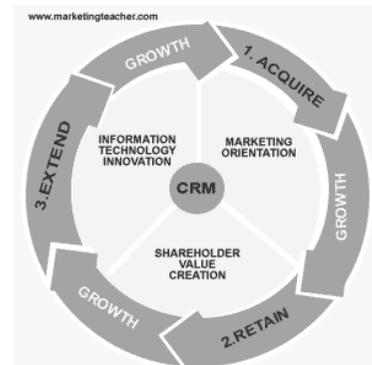

Figure 1. Business Strategy and CRM Model

The customer acquisition is the process of attracting customer for the first of their purchase or use services. The customer retention is the customer return to us and uses the service for the second time. We keep them as customer. And the customer extension is introducing new product or service line to our loyal customers that may not relate to the original service or product. Growth the numbers of new or retain customers use the product or service through marketing orientation, value creation and innovative IT.

Web 2.0, which play a significant part in the CRM transition drives social change that impacts all institution including business and healthcare organizations. It is a revolution on how people communicate. It facilitates peer-to-peer collaboration and easy access to real time communication and that is core of social change. Because much of the communication transition is organized around web based technologies, it is called Web 2.0 (P. Greenberg, 2009). Patients participate in these social network can share information about their diagnoses, medications, healthcare experiences, and other information. It is often in form of unstructured communication which can provide new insights for people involved in the management of health status and chronic care conditions.

Social CRM is based on the Web 2.0. The Web 2.0 could be used as enablers in creating close and long term relationships between an organization with its customers (Askool, and Nakata, 2010). The concept of Web 2.0 began with a conference brainstorming session between O'Reilly and MediaLive International (O'Reilly, 2005). It has been defined as a set of economic, social, and technology trend that collectively form the basis for next generation of the Internet – a distinctive medium characterized by user participation, openness network effects (O'Reilly, 2006). Recently, the Web 2.0 tools such as Facebook, Twitter, Myspace, Friendster, LinkedIn, etc. have grown rapidly facilitating peer-to-peer collaboration, ease of participation, and ease of networking. However, the effects of Web 2.0, particularly in addressing the issues of customer relationship have not been explored. As the main advantages of Web 2.0 are the linkage among people, ideas, processes, systems, contents and other organizational activities (Askool, and Nakata, 2010). Therefore, Web 2.0 definitely will affect performance of the organization as it is about engaging relationships, sharing experience & information, and collaboration.

Greenberg (2009) defined Social CRM as a philosophy and a business strategy, supported by a technology platform, business rules, processes, and social characteristics, designed to engage the customer in a collaborative conversation in order to provide mutually beneficial value in a trusted and transparent business environment. It's the company's response to the customer's ownership of the conversation.

The term of Social CRM and CRM 2.0 is used interchangeably. Both share new special capabilities of social media and social networks that provide powerful new approaches to surpass traditional CRM.

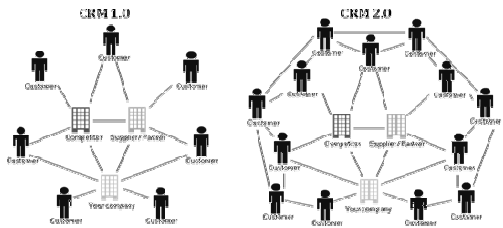

Figure 2. Evolution of CRM landscape (Fabio Capriani)

Fabio Cipriani described the fundamental changes that Social CRM is introducing to the current, traditional CRM in term of landscape. Figure 3 is reflection of the evolving CRM 2.0 which is different from CRM 1.0. It is a revolution in how people communicate, customers establish conversation not only with the service provider but it is also with others. Table 1 summarizes the difference of CRM 2.0 from CRM 1.0 based on type of relationship, connection, and how value generated. Relationship type in CRM 1.0 focuses on the individual relationship; Customer to Customer or Customer to Business whereas in CRM 2.0 offers the collaborative relationship and engage a more complex relationship network. Connection type in CRM 1.0 is limited view of the customer which affect to less informed customer, on the other hand, CRM 2.0 enable for multiple connections allow better understanding and more knowledgeable customer. CRM 1.0 of value creation is constricted from targeted messages, and CRM 2.0 offers diverse value creation even from informal conversation of customers within social networks.

Social networking could generate a way to strengthen relationship between organization and their customers. The Web 2.0 is an important tool for the development of social network. In addition, Web 2.0 which plays a significant part in the CRM transition, stimulated fundamental changes in consumer behavior (Greenberg, 2009). This revolution is having a broad and deep impact on an interpersonal relationship in all areas, and health is no exception. The booming number of social networking groups and supports groups for patients on the internet and their influence on health behavior is only beginning to be explored (Rimer and others, 2005) and remains an important area for future research. The concept of a social network defines organizations as a system that contains objects such as people, groups, and other organizations linked together by a range of relationships (Askool, and Nakata, 2010).

III. DISCUSSION

The gap between existing CRM systems and customer care needs make it more complex. CRM can be viewed as strategy to attract new customers coming to an organization, retaining them throughout the entire lifetime of a relationship, and extending other services or products to the existing customers. In the healthcare environment, healthcare organizations are challenged to acquire potential customers for the healthcare services, retaining them to use the services, and extending various services in the future. To take the challenges, healthcare organization must consider establishing close of relationship with their patients offer convenience of services, and provide transparency in services through information sharing. Therefore, the healthcare organization should perform re-engineering process to adapt their CRM strategy and tool in order to acquire potential customer coming for the service (Anshari and Almunawar, 2011).

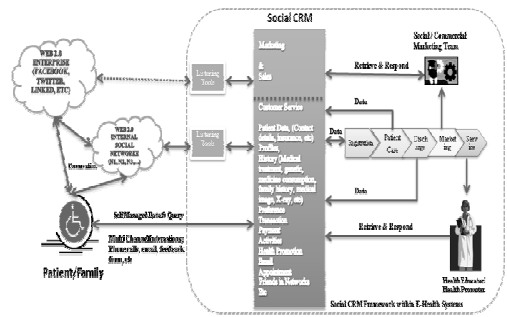

Figure 3. Model of CRM 2.0 in E-Health Services

Figure 3 shows the components of CRM 2.0 in the e-health scenario. CRM 2.0 promotes openness which all activities involve with patient recorded on systems and patients/families are able to access them online. The service is more or less the same with the traditional healthcare system; however the different is the privileges to access medical records, patient personal data, appointment with physician, scheduling, and any other features of Web 2.0 added to establish conversation, convenience, and creating trust to the service is innermost for healthcare that employ CRM 2.0.

The model differentiates two social networks linkages to the patient; they are Enterprises Social Networks and Internal Social Networks. The Enterprises Social Networks in this framework refers to external and popular Web 2.0 applications such as *Facebook*, *Twitter*, *LinkedIn*, *MySpace*, etc which patient may belong to any of those social networks for interaction. The dashed line connected enterprises social networks and CRM systems mean that none of those networks have control over the others directly, but constructive conversation and information from enterprises social networks should be captured for creating strategy, innovation, better service and at the same time responded accurately. Lesson learnt from Prita's case, the hospital was not proficient capturing the message from the customers at social networks because they did not consider in their CRM strategy that the customers have changed and they made conversation, judging hospital's value, criticizing their services at those networks which led to distrust towards the hospital's allegation and jeopardized the business for the long run.

Additionally, the framework proposes Internal Social Networks that operated, managed, and maintained within healthcare's infrastructure. This is more targeted to internal patients/families within the healthcare to have conversation patient with the same interest or health problem/ illness. For example, patient with diabetic would motivate to share his/her experiences, learning, and knowledge with other diabetic patients. Since patient/family who generates the contents of the Web, it can promote useful learning center for others, not only promoting health among each others, but also it could be the best place supporting each other and sharing their experiences related to all issues such as; how the healthcare does a treatment, how much it will cost them, what insurance accepted by healthcare, how is the food and nutrition provided, etc. Therefore, this is generic group that will grow depends on the need of patients in that healthcare. For instant, N1 is internal social networks for Diabetic, N2 is for Cancer, N3 is for hearth disease, and so on. Creating Internal Social Networks is part of the strategy to isolating problem into small space or more focus to the local's problem so it can be easily monitored and solved. Moreover, this strategy will promote loyalty of customers to keep using service from the healthcare.

The absence of this strategy in the Prita's case, the hospital was late to realize that the patient dissatisfied with the service from beginning, and the hospital assumed that everything was fine, until she communicated her dissatisfaction through her social networks. Responding this problem, the hospital should isolate the internal problem like dissatisfaction of patient by quick response to resolve the issue before it gets bigger and uncontrolled. And the Internal Social Networks could be solution to prevent the same problem in the future.

In general, the aim to put together linkage of internal and external social networks are to engage patients and export ideas, foster innovations of new services, quick response/feedback for existing service, and technologies from people inside and outside organization. Both provide a range of roles for patient or his/her family. The relationships can create emotional support, substantial aid and service, influence, advice, and information that a person can use to deal with a problem. In addition, listening tool between Social Networks and CRM systems (see figure 6.) is mechanism to capture actual data from social media and propagates this information forward to the CRM. This tool should be capable to filter noise from actual data that needs to be communicated to CRM.

The foundation of CRM 2.0 which is based on Web 2.0, it empowers patient/family to have ability controlling their own data. Once patient/family registers to have service from healthcare, it will enable them to have personalized e-health systems with Social CRM as frontline of the system. The system will create account for each patient then; the authorization and self-managed account/service are granted to access all applications and data offered by the systems. This authorization is expected to be in the long run since the information and contents continue to grow. Technical assistant is available through manual or health informatics officer (just like any other customer service in business/organization) who stand by online assisting patient/family in utilizing the systems especially for the first timer. Furthermore, since all the information (medical records) can be accessed online everywhere and anytime, it will enable collaborative treatment from telemedicine.

CRM 2.0 functionalities compose from Marketing, Sales, and Customer Service. The different from the traditional CRM, the state to empowering for self managed data and authorization will encourage patient willingly to provide full data without hesitation. More data provided more information available for the sake of analyzing for the interest of marketing, sales, and customer service. Suppose this scenario; we go to Physician for diagnose, and the doctor is well known physician in town with long queue patients waited, once we get the chance for diagnose, how long he will have time to investigate the symptom of the problems? The system will improve the customer service because it helps both parties either physician or patient in diagnose activity. The doctor will have complete information, knowledge, and saving a lot of time to learn about patient history because patient participate in the detailing his medical records data through the system, and patient benefits from quality of diagnoses' time because his medical records are overviewed in full scene.

The other feature of the model is robustness of systems because more applications/services will be added as characteristics of Web 2.0. Some of the features that available to the user are; updating personal data, Medical Records & History (medical treatment received, medicine consumption history, family illness history, genetic, medical imaging, x-ray, etc), Preference services, Transaction, Payment/Billing data, Activities, Personal Health Promotion and Education, Email, Appointment, Friend in networks, forums, chatting, etc.

One of the goals CRM 2.0 in healthcare is providing value-added services to patients like openness of medical records, improving patient loyalty, creating better healthcare-patient communication, improving brand image and recognition, and self managed data which will improve health literacy to reduce economic burden for society to the whole. The raw data material arrives in one state, and leave in another state. The patient enters ill and leaves well. The activities of value creation in healthcare are; arriving from registration, patient care, discharge, marketing, and service—producing data at respective state. The own unique characteristics value creation by adopting CRM 2.0; generate contents from both parties either from healthcare and also patients.

The other features of CRM 2.0 in healthcare come into view in respond to better customer service. By empowering patients with medical data and personalized e-health needs healthcare to provide health educator or health promoter to interpret medical data to easily understood by patient/family or in respond to online query/consultation. Officer in duty is required to have an ability to interpret medical data and also familiar with the technical details of the systems. Another instance from the framework is Social/Commercial Marketing team. Social marketing is more prevalence to the government healthcare that operates as an agent of the public at large. On the other hand, commercial marketing is standard marketing strategy exist for any business entities. Both are acting in responds to the public demands like social networks, Mailing list, etc. The adoption of Social CRM to healthcare prevents any dispute and avoiding conflict between healthcare and patient. Prita's case in introduction took place because; the hospital needs to understand that behavior and expectation of patients continue to change eventually. And this study proposes that CRM 2.0 framework as alternative solution to the hospital. Once the case became a public knowledge, it affects survival ability of the hospital in the long run jeopardize due to loosing of the trust towards the service. Therefore, the hospital should perform re-engineering process to adapt their CRM strategy and tool in order to acquire potential customer coming for the service.

IV. CONCLUSION

The complexity of healthcare's business process manages relationship between patient and healthcare is one of the most remarkable aspects in medical process. In fact, patient expectations in healthcare are high, which need trust in delivering service. A new paradigm has appeared in CRM systems namely Social CRM or CRM 2.0 as a result of the development of Web 2.0 technology. By inheriting features from the Web 2.0 technology, CRM 2.0 offers new outlook either from patient or healthcare. Some of the features offered by Social CRM framework are robustness of the systems, trust of information sharing, and closeness of relationship between patient-healthcare and patient with others. The systems create value in each activity to the customer. And those values will make the healthcare a better service than its competitors. Moreover, it empowers patients with the data accessibility in returns of loyalty and trust relationship with their patients.